\documentclass[a4paper,11pt]{article}
\usepackage{pos}
\usepackage{comment}
\usepackage{xcolor}
\usepackage{lipsum}
\usepackage{subcaption}
\usepackage{setspace}
\usepackage{sidecap}
\usepackage{wrapfig}
\usepackage[numbers,sort&compress]{natbib}

\title{The Trinity-One PeV-Neutrino Telescope }

\author*[a]{David A.~Raudales O.}
\author[a]{A.~Nepomuk Otte}
\onbehalf{for the \emph{Trinity} Collaboration}

\affiliation[a]{School of Physics and Center for Relativistic Astrophysics, Georgia Institute of Technology,\\
  Atlanta, Georgia, USA}

\emailAdd{draudales@gatech.edu}

\abstract{Following the Trinity Demonstrator, Trinity One will be the first of the 18 Cherenkov telescopes that make up the Trinity PeV-Neutrino Observatory. Located on Frisco Peak in Utah, Trinity One can observe 64\% of the sky, allowing it to detect potential neutrino point sources with unprecedented sensitivity, ranging from 1\,PeV to 10\,EeV. We outline the design of Trinity One, which features a 60\,m$^2$ light-collection surface and the ability to rotate in azimuth. It has a field of view measuring $5^\circ$ by $60^\circ$, which is equipped with a silicon photomultiplier camera with a resolution of $0.3^\circ$. Utilizing the design of Trinity One, we present performance calculations in relation to various source classes.}

\ConferenceLogo{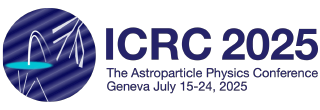}

\FullConference{39th International Cosmic Ray Conference (ICRC2025)\\
 15–24 July 2025\\
Geneva, Switzerland\\}

\begin{document}
\maketitle

\section{Introduction}

Neutrinos have proven to be versatile tools in exploring some of the most fascinating facets of the non-thermal ultra-relativistic universe. The IceCube team has laid the groundwork for high-energy neutrino astronomy with an extraordinary collection of neutrino measurements ranging from the detection of our own Milky Way in neutrinos \cite{Abbasi2023a}, the astrophysical diffuse neutrino flux \cite{Aartsen2017}, to the potential discovery of neutrino emission from the vicinity of two extragalactic supermassive black holes \cite{Abbasi2022,Aartsen2018b}.   

These measurements at high energies (1\,TeV to 1\,PeV) demonstrate the need for instruments with not only significantly improved sensitivity at high energies but also with adequate sensitivity at very-high energies (VHE, 1\,PeV to 1\,EeV). By opening the VHE band, we gain access to the tail of the astrophysical neutrino flux \cite{Aartsen2017}, we can probe the predicted neutrino emission from blazars \cite{Petropoulou:2019zqp} and gamma-ray bursts \cite{Lian:2024mxq}, and we can fully explore cosmogenic neutrinos, \emph{i.e.} neutrinos produced in interactions of $>$EeV cosmic rays with the CMB \cite{Aloisio2015}, and their implications for the evolution of cosmic ray sources and cosmic ray composition.

Of the different avenues currently pursued by the neutrino astroparticle community to gain sensitivity in the VHE neutrino band \cite{Ackermann2022}, we pursue the development of the \emph{Trinity} Neutrino Observatory. \emph{Trinity} uniquely bridges the lower VHE regime, overlapping with IceCube, while opening access to a vast region of the VHE sky inaccessible to IceCube.

In the next section, we provide an overview and status of the \emph{Trinity} development. The remainder of the paper discusses the design of \emph{Trinity} One and \emph{Trinity} One's potential to detect and constrain VHE neutrino emission from GRB's and Blazars based on model predictions.

\section{The \emph{Trinity} Neutrino Observatory}

The \emph{Trinity} Neutrino Observatory concept envisions the deployment of several wide-angle Cherenkov telescopes with an effective light collection area of more than $ 10$\,m$^2$ on mountain tops. Pointing at the horizon, these telescopes can detect the Cherenkov light emitted from air showers from 200\,km away, yielding the necessary large acceptance to detect air-showers initiated by Earth-skimming VHE tau neutrinos \cite{Otte2019d}. These neutrinos enter the Earth under a shallow angle of $<10^\circ$, yielding a high probability to generate a tau in a charged current interaction inside the Earth, and the tau then emerges into the atmosphere where it decays and starts an air-shower \cite{Fargion2006,Ahnen2018a,Ogawa2019,Otte2019d}.

\begin{SCfigure}[0.7][!tb]
\centering
\includegraphics[trim={2cm 5cm 5cm 5.0cm},clip,angle=0,width=0.7\columnwidth]{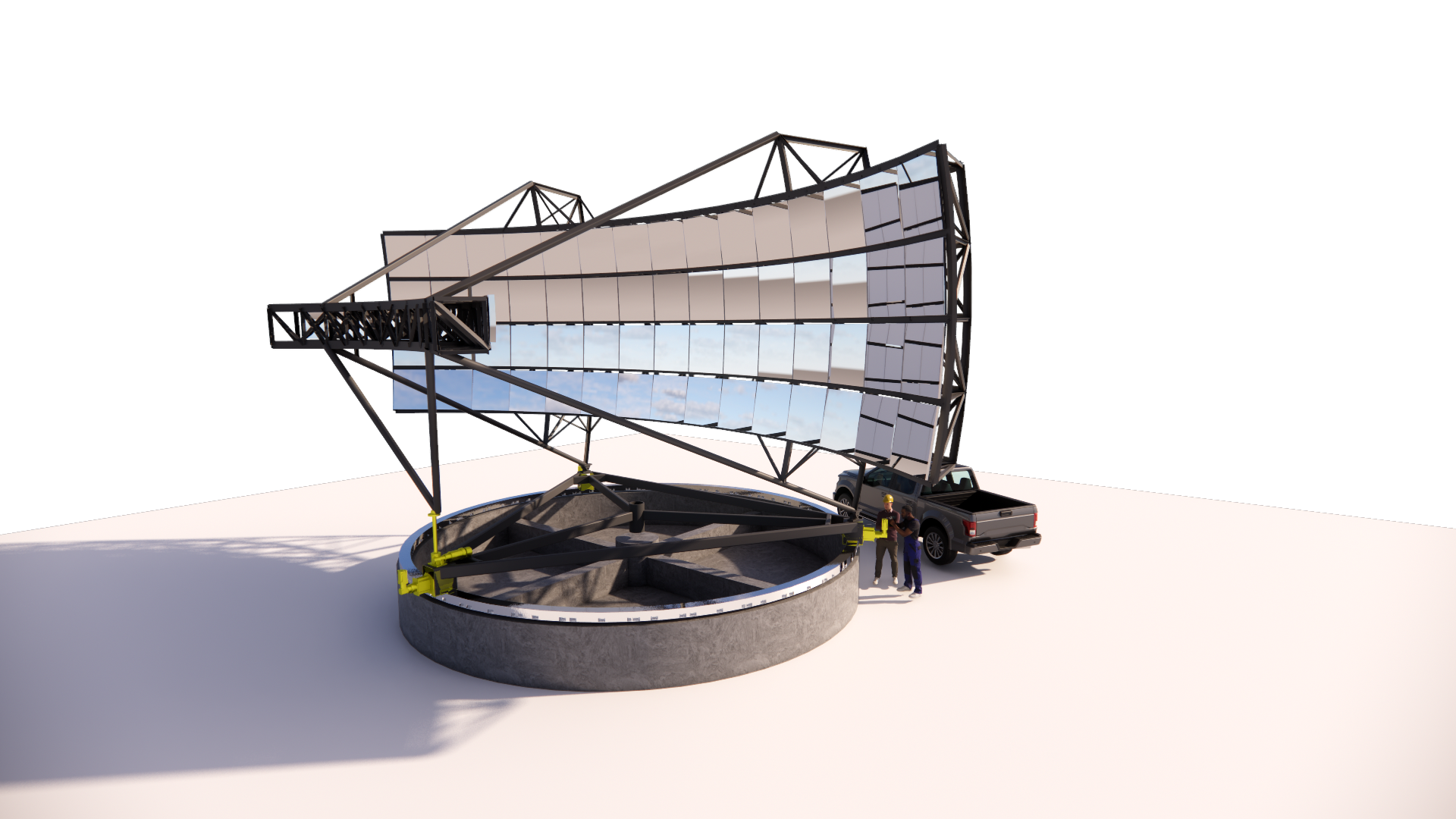}
\caption{Rendering of the \emph{Trinity} One telescope.}
\label{fig:TrinityOne}
\end{SCfigure}

\emph{Trinity} is developed in three phases. The \emph{Trinity} Demonstrator, followed by \emph{Trinity} One, and subsequently scaling to the full \emph{Trinity} Neutrino Observatory. Deployed in Fall 2023, the \emph{Trinity} Demonstrator successfully completed commissioning in Summer 2024. Given the encouraging results obtained with the Demonstrator\footnote{Results from 346 hours of observations with the Demonstrator are presented elsewhere in these proceedings.} \cite{Stepanoff_2025,Bagheri:2025fxh}, we are ready to develop \emph{Trinity} One.

\section{\emph{Trinity} One}

\emph{Trinity} One will be the first telescope of the \emph{Trinity} Observatory. Figure \ref{fig:TrinityOne} shows a CAD rendering of the telescope. \emph{Trinity} One has a $60^\circ$ horizontal field of view and $5^\circ$ in the vertical. Its segmented 60\,m$^2$ light collection surface provides an effective 16\,m$^2$ light collection area for each of the 3,328 silicon photomultiplier pixels of the camera. With that pixelization, the camera yields a $0.3^\circ$ angular resolution. The camera is read out with a 250\,MS/s sampling rate and a 12-bit resolution. \emph{Trinity} One can rotate in azimuth and, therefore, it can observe sources over a large range of declinations, when deployed on Frisco Peak, Utah. The annual acceptance for any point in the sky, accounting for its 20\% duty cycle and source visibility, is shown in Figure \ref{fig:three_skymaps}.

\section{\emph{Trinity} One's Effective Area, Acceptance and Fluence sensitivity}

\begin{SCfigure}[0.7][htb]
    \centering
\includegraphics[width=0.7\linewidth]{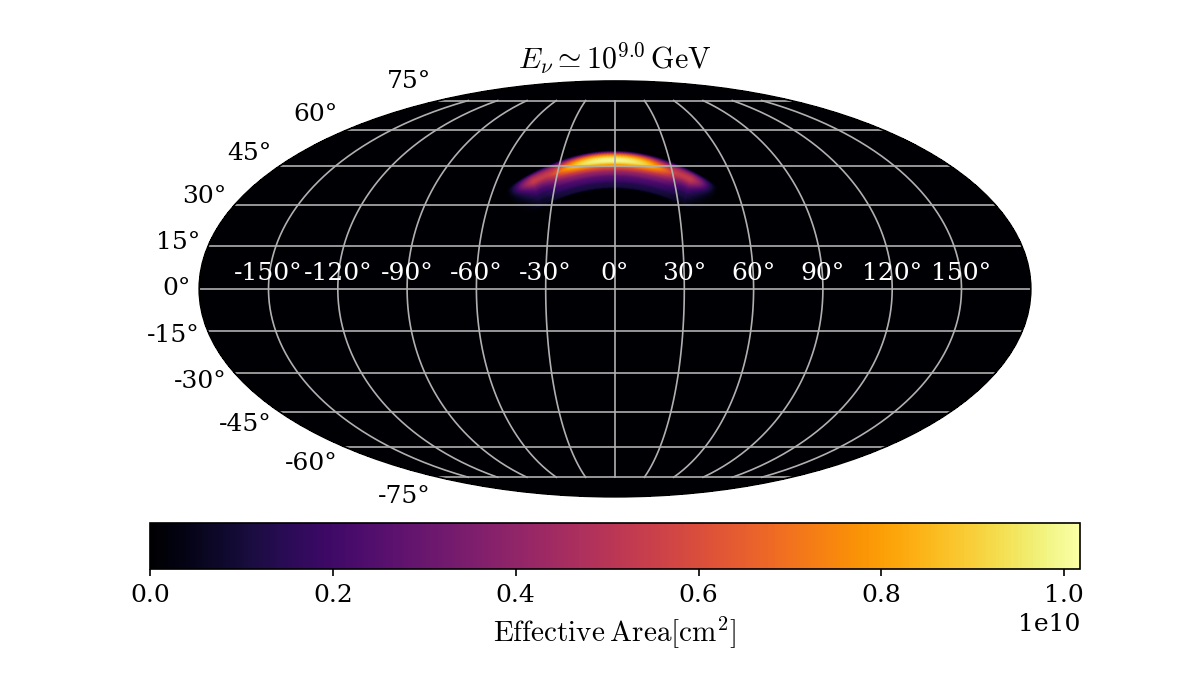}
    \caption{Instantaneous effective area to $10^{18}$ eV tau neutrinos for one Trinity Telescope located in Frisco Peak, Utah, United States (Local Sidereal Time of 0:00)}
    \label{fig:instEA}
\end{SCfigure} 

The calculation of Trinity's effective area takes into account the probabilities that a) a neutrino with a given energy and direction interacts inside the Earth, and a tau emerges from the ground, b) the air shower following the tau decay fully develops inside Trinity's field of view of Trinity, c) the telescope receives enough light to trigger the readout and later reconstruct the event, and d) the length of the air shower is more than $0.3^\circ$. For details, see \cite{Otte2019d, Wang_2021}. 
The instantaneous effective area, $A_{eff}\bigl(\phi,\theta_z,E_\nu)$, for a single Trinity telescope is shown in Figure \ref{fig:instEA} for a neutrino energy of $E_\nu = 10^9$\,GeV. 
To study \emph{Trinity} One's  response to point sources at different timescales, we calculate the integral acceptance by weighting it with a power law source spectrum  $E_\nu^{-2}$,
\begin{equation}
    \mathcal{A}\bigl(\phi,\theta_z) = \dfrac{1}{N}\int dt \int dE_{\nu}A_{eff}\bigl(\phi,\theta_z,E_\nu) E_\nu^{-2}
\end{equation}
where $N =  \int dE_{\nu}E_\nu^{-2}$ is the normalization. 

\begin{wrapfigure}[32]{r}{0.6\textwidth}
    \vspace{-5.5ex}\includegraphics[width=\linewidth]{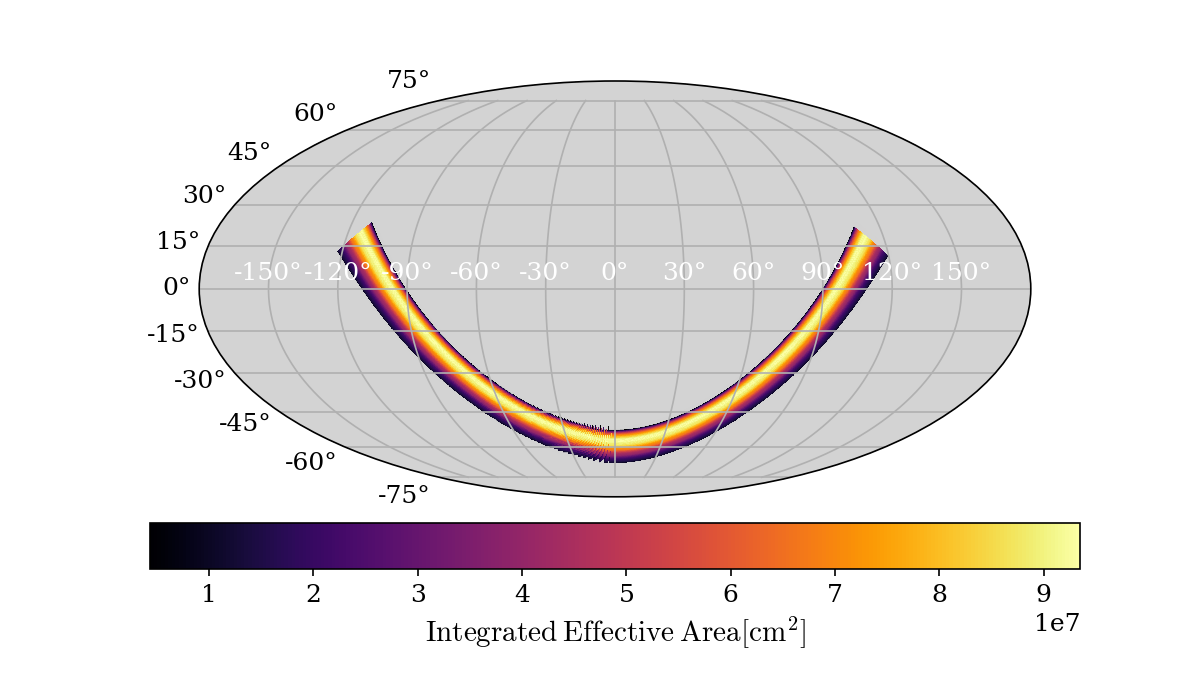}
    \begin{center}
      \includegraphics[width=0.85\linewidth,clip,trim=0.5 0.5 0.5 0.5]{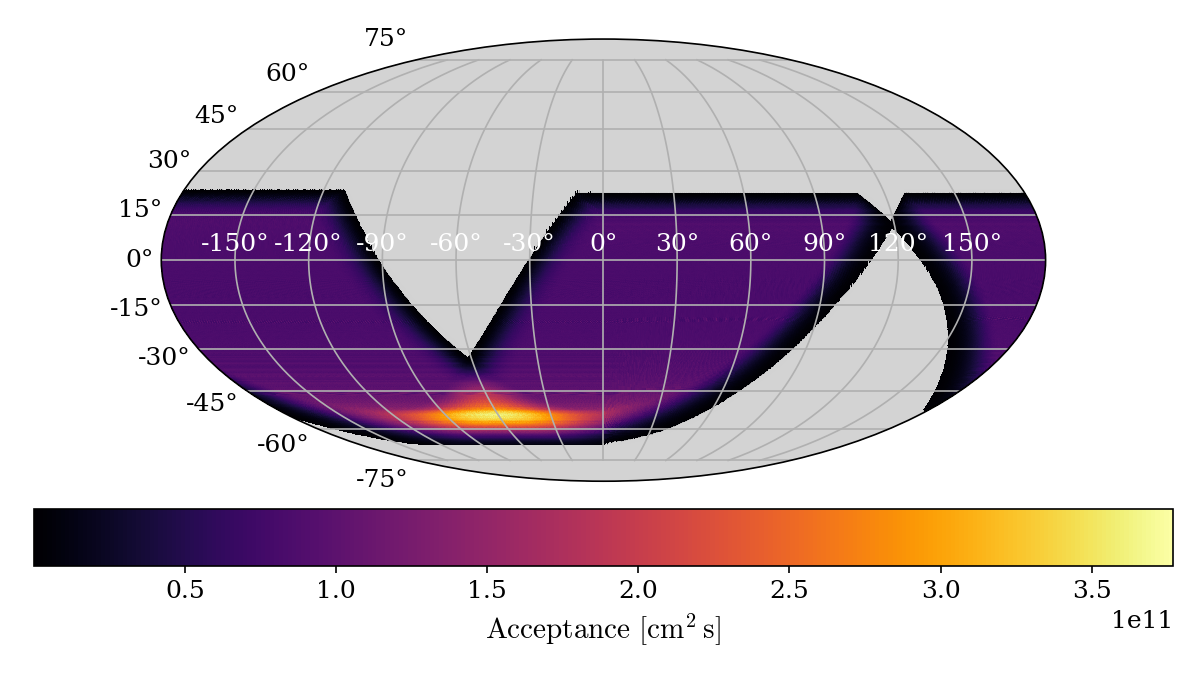}
    \end{center}
    \includegraphics[width=\linewidth]{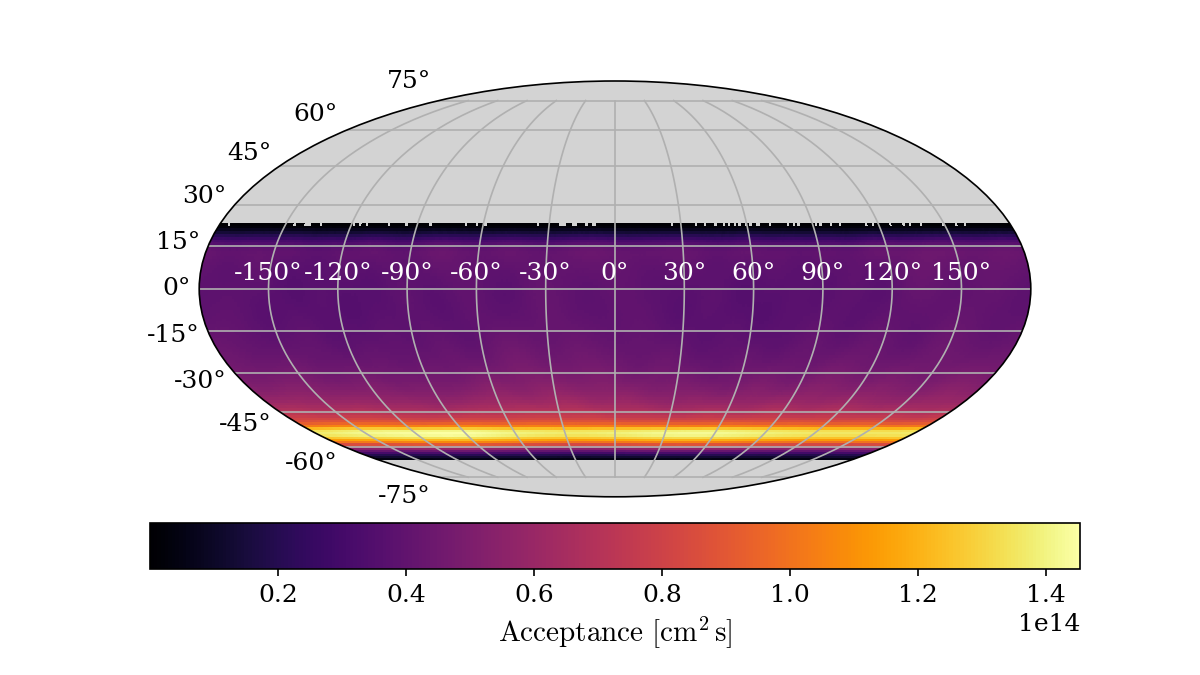}

  \caption{Top: Instantaneous effective area, weighted by an $E_\nu^{-2}$ spectrum, considering azimuthal rotations from 60° to 300°, Middle: Resulting acceptance after an eight-hour night of observations, Bottom: Acceptance integrated over one year.}
  \label{fig:three_skymaps}
\end{wrapfigure}
The instantaneous effective area above does not take into consideration that \emph{Trinity} One can rotate in azimuth between $60^\circ$ and $300^\circ$ on Frisco Peak and point at different sources. We, therefore, introduce the \emph{accessible} integral effective area, which quantifies the effective area of \emph{Trinity} One if it were slewed to the corresponding azimuth. The top panel in Figure \ref{fig:three_skymaps} shows the thus calculated \emph{accessible} instantaneous integral effective area for a random time. Correspondingly, the \emph{accessible} acceptance quantifies the possible acceptance for any point in the sky, \emph{i.e.}, it is useful to quickly identify if a source is observable and with what possible acceptance considering re-pointing of the telescope. The middle panel in Figure \ref{fig:three_skymaps} shows the \emph{accessible} acceptance for one eight-hour night. And the bottom panel shows the \emph{accessible} acceptance for one full year of observing. For the annual acceptance, we take into account that we can only observe on sufficiently dark nights, resulting in a 20\% duty cycle.

\begin{SCfigure}[0.7][htb]
    \centering
\includegraphics[width=0.7\textwidth]{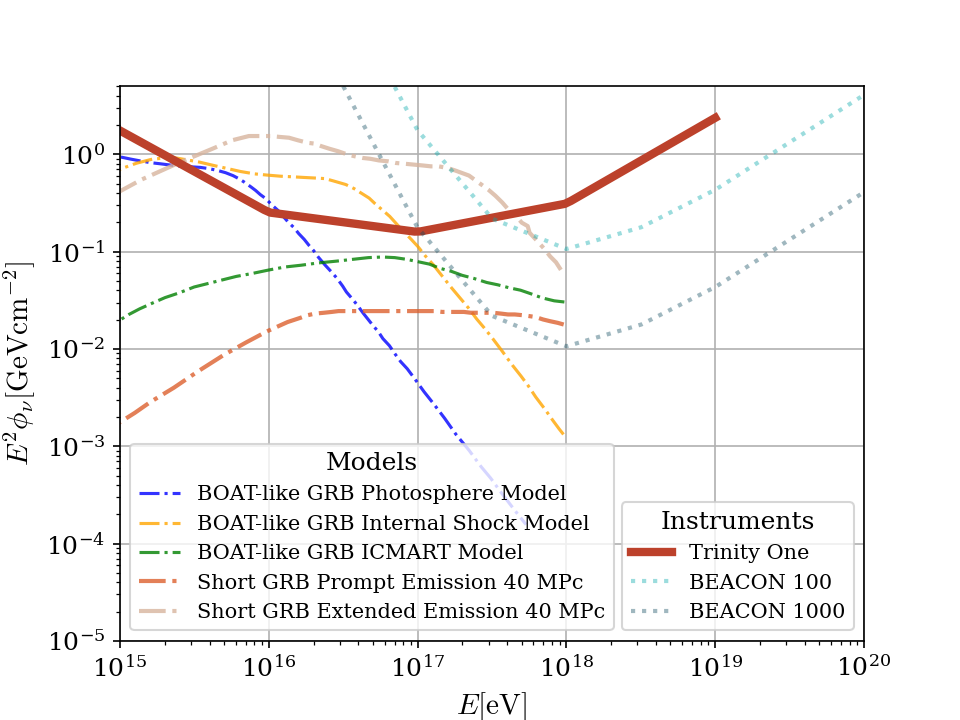}
    \caption{Short-Burst (< 1000 s) Fluence Sensitivity of \emph{Trinity} One compared to several GRB models \cite{Lian:2024mxq,Kimura:2017kan} and BEACON's fluence sensitivity. \cite{BEACON:2025qcq}}
    \label{fig:sensitivity}
\end{SCfigure}

Based on \emph{Trinity} One's maximum instantaneous effective area for each energy decade and the assumption of an all-flavor neutrino power-law spectrum ($E_\nu^{-2}$), \emph{Trinity} One's maximum fluence sensitivity for short transients ($<1,000$\,s) is shown in  Figure \ref{fig:sensitivity} calculated with the following fluence sensitivity definition:

\begin{equation}\label{eq:sensitivity}
E^2_\nu S(E_\nu) = \frac{2.44}{\Delta(\log_{10}E_\nu)} \frac{3}{\ln(10)} \frac{E_\nu}{A_{\text{eff}}(E_\nu)}
\end{equation}

where 2.44 represents the Feldman-Cousins upper limit per energy decade at a 90\% confidence level (C.L.) for a background-free observation. For a discussion of diffuse flux sensitivity, see \cite{Stepanoff_2025}. Implications for the observation of transients with \emph{Trinity} One are discussed in the next section.

\section{Source Observations with \emph{Trinity} One}

Detecting sources remains one of the central objectives in neutrino astronomy and will be a strength for \emph{Trinity} One. To exemplify \emph{Trinity} One's potential, we discuss its ability to detect PeV neutrinos from a flaring blazar, high-luminosity GRBs, and short GRBs.

\paragraph{Flaring Blazars:} Lacking a firm detection of PeV neutrino emission from a blazar, we adopt the evidence of neutrino emission observed with IceCube from the direction of TXS 0506+056 during 158 days as a baseline flux and investigate how many neutrinos \emph{Trinity} One would detect from such a source at different distances and flaring times. 

\begin{SCfigure}[0.7][htb]
    \centering    \includegraphics[width=0.6\textwidth]{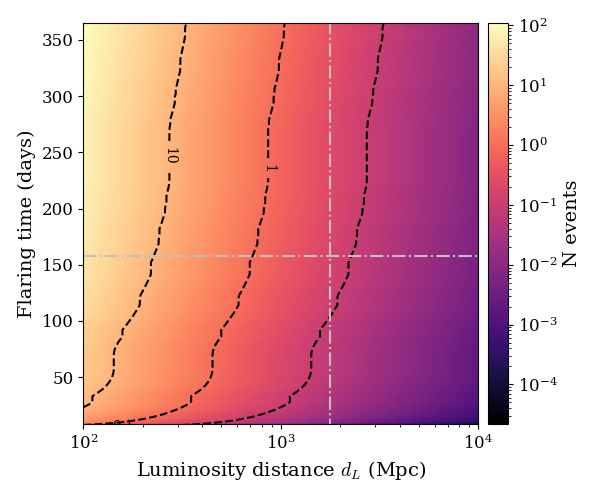}
    \caption{Expected number of neutrino detections with \emph{Trinity} One for a TXS 0506+056 like source at distances up to 10\,Gpc and flaring times up to one year. The grey dashed lines mark the distance of TXS 0506+056 and the 158 day flaring period identified by IceCube.}
    \label{fig:blazar}
\end{SCfigure}

Because the spectral shape of the neutrino spectrum is not fully constrained, we assume the neutrinos are produced in photohadronic interactions, yielding the spectral energy distribution in \cite{Petropoulou:2019zqp} normalized to the IceCube data, see Figure 4 of \cite{Petropoulou:2019zqp}. With that \emph{standard candle} flux, we then calculate the number of neutrinos detected with \emph{Trinity} One for different flaring times and source distances with
\begin{equation}
    \label{expected_events}
    N_\nu = T_{obs}\int dE_\nu F_\nu (E_\nu) \langle A_{eff} (E_\nu)\rangle 
\end{equation}
where the observation time, $T_{\rm obs}$, is defined as the total duration during which a source located at the same sky position as TXS~0506+056 is visible to \emph{Trinity One}. For the 158-day flaring period, this corresponds to 116 hours of exposure. To account for the variation in sensitivity as the source traverses \emph{Trinity One}'s field of view, we employ the direction-averaged effective area of the instrument. The resulting expectations are shown in Figure~\ref{fig:blazar}, which displays the number of detected events as a function of source distance (up to 10\,Gpc) and flare duration (up to one year). 

For TXS~0506+056 during its 158-day 2014-2015 flare, \emph{Trinity One} would have expected to detect $0.16$ neutrinos, corresponding to a Poisson probability of $15\%$ for at least one event. For comparison, a similar flare from a source at $1\,\mathrm{Gpc}$ would yield approximately one detected neutrino within one year of observation.

\begin{figure}[htbp]
  \centering
  \begin{minipage}[b]{0.48\textwidth}
    \includegraphics[width=\linewidth]{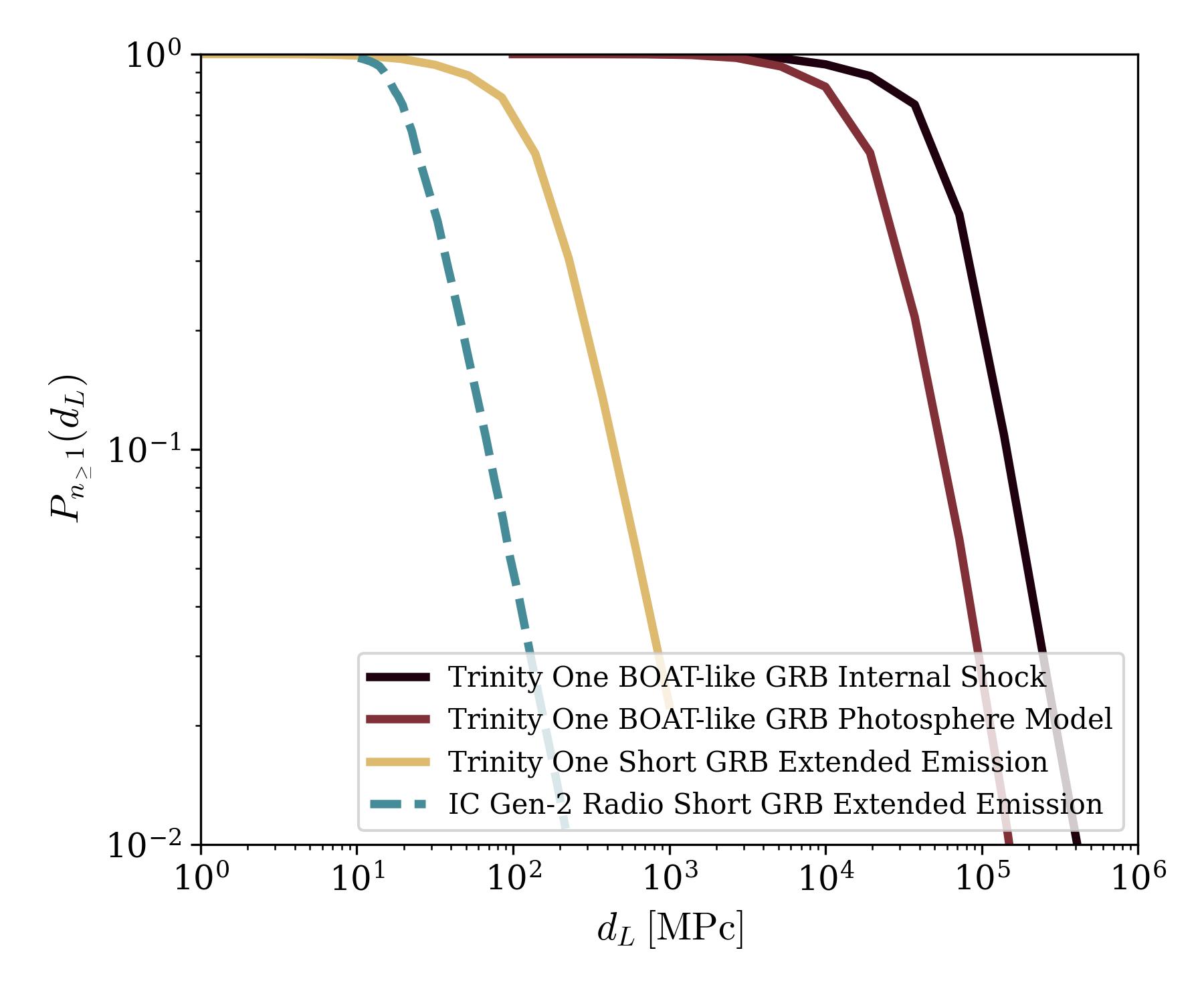}
  \end{minipage}\hfill
  \begin{minipage}[b]{0.48\textwidth}
    \includegraphics[width=\linewidth]{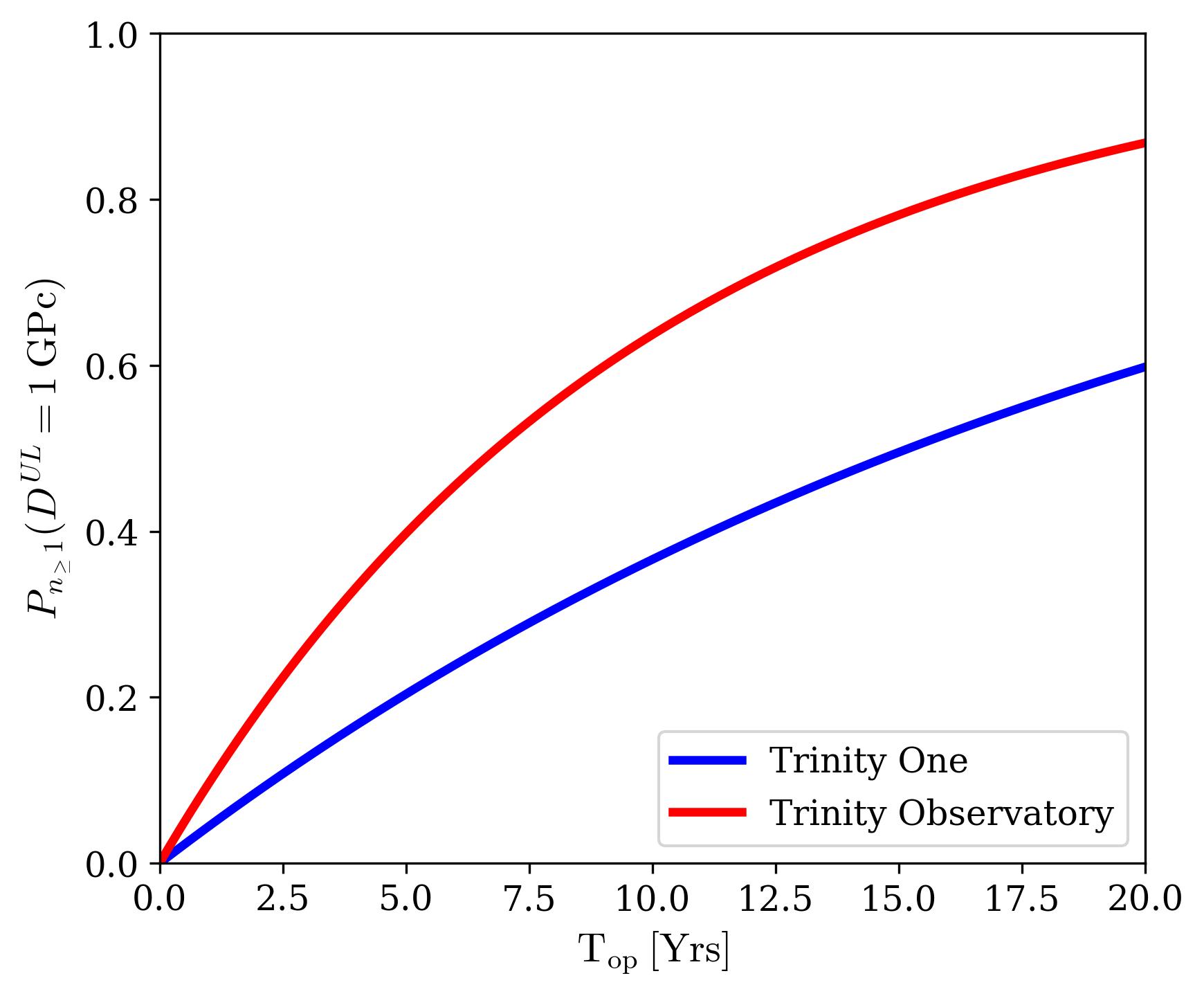}
  \end{minipage}
  \caption{Left: Probability of detecting at least one neutrino from short GRBs or BOAT-like GRBs vs.\ luminosity distance with \emph{Trinity} One; Right: Probability of detecting at least one neutrino from a short GRB as a function of operation time with \emph{Trinity} One and projected \emph{Trinity} Observatory.}
  \label{fig:GRBprobabilities}
\end{figure}

\paragraph{High-Luminosity Gamma-Ray Bursts (HLGRBs)} GRBs represent prime candidates for PeV neutrino detection with \emph{Trinity} One, owing to the overlap between their predicted emission spectra and \emph{Trinity} One’s energy coverage. A non-detection could potentially rule out GRB models. Here we explore the probability to detect neutrinos from a GRB 221009 A (the BOAT) like event. We explore Trinity One's detection capability of such an event following the formalism presented in \cite{Kotera:2025jca}. The probability of detecting at least one neutrino event from a given source at luminosity distance $d_L$ is given by:

\begin{equation}
P_{n \geq 1}(d_L) = \frac{1}{\Omega_{\text{norm}}} \int_{\Omega} d\Omega  p_{n \geq 1}(\phi, \theta, d_L)
\end{equation}

where $ p_{n \geq 1}$ represents the Poisson probability of detecting at least one neutrino:

\begin{equation}
p_{n \geq 1}(\phi, \theta, d_L) = 1 - \exp\left(-N_\nu(\phi, \theta, d_L)\right)
\end{equation}

averaged over the detector's solid angle \cite{Mukhopadhyay:2024lwq}. The number of neutrino events is calculated for each point in the field of view for two different models using Eq. \ref{expected_events}. The left panel in Figure \ref{fig:GRBprobabilities} shows that \emph{Trinity} One would detect at least one neutrino if a BOAT-like event happens within 10\,Gpc and goes off in its field of view. 

\paragraph{Short Gamma-Ray Bursts (sGRBs)} 
sGRBs, widely believed to originate from the merger of neutron stars, are uniquely powerful multi-messenger sources, observable in both electromagnetic and gravitational-wave channels. This makes them prime targets for follow-up studies with \emph{Trinity} One. With the upcoming Einstein Telescope and Cosmic Explorer expected to detect thousands of binary neutron star mergers per year \cite{Ronchini:2022gwk}, the prospects for joint observations are particularly compelling. Following the methodology outlined above, we compute the probability of detecting at least one neutrino from an sGRB occurring within \emph{Trinity} One's field of view (left panel of Figure~\ref{fig:GRBprobabilities}). The right panel illustrates the cumulative probability of an sGRB both occurring within the accessible sky region and it being detected with at least one neutrino by \emph{Trinity}. For a single telescope, this probability corresponds to 60\% after twenty years of operation. To provide an estimate for the full Trinity Observatory, we assume a daily sky coverage of about 80\%, corresponding to six telescopes distributed across three sites. Moreover, we use the same direction-averaged effective area as the calculated for \emph{Trinity} One. Under this assumptions, the probability of detecting at least one neutrino event after twenty years approaches 90\%.

\section{Discussion}
The \emph{Trinity} Neutrino Observatory targets the PeV to EeV energy range and is developed in three stages. Since Fall 2023, we have operated the \emph{Trinity} Demonstrator, the first stage towards the \emph{Trinity} Observatory. The purpose of the Demonstrator was to identify potential sources of background and develop relevant technologies and analysis tools. Both objectives have been met, which is why we are transitioning towards \emph{Trinity} One, the first telescope of the \emph{Trinity} Observatory. \emph{Trinity} One will have a $60^\circ$ horizontal field of view, which results in a superb point source sensitivity in the PeV to EeV energy range for a fraction of the cost of other proposed and operating instruments. Because \emph{Trinity} One will rotate in azimuth, it is sensitive to neutrinos from sources in a broad range of declinations between $-67^\circ$ and $23^\circ$.

Adopting the evidence for neutrino emission from TXS~0506+056 as a \emph{standard candle}, we find that \emph{Trinity} One would have expected $0.16$ PeV–EeV neutrinos during the 158-day 2014–2015 flare, corresponding to a $15\%$ probability of detecting at least one event. The detection probability increases for a closer source or a longer flare, even after accounting for the instrument’s $20\%$ duty cycle.

In addition to blazars, \emph{Trinity} One is sensitive to neutrino emission from short and high-luminosity GRBs. A sGRB within $100\: \mathrm{Mpc}$ would be detectable, while a BOAT-like burst could be observed out to $10 \:\mathrm{Gpc}$, provided it occurs within the field of view. For sGRBs, the cumulative probability of at least one detection with a single telescope reaches $60\%$ over $20$ years; extrapolating to the full \emph{Trinity} Observatory, six telescopes at three sites with $80\%$ daily sky coverage, the $20$-year probability approaches $90\%$.

As the inaugural element of the planned array, \emph{Trinity} One will deliver valuable insights on neutrino emission from the most energetic transient phenomena in the universe, while serving as a pathfinder that lays the experimental and scientific groundwork for the broader impact of the full \emph{Trinity} Neutrino Observatory.

\begingroup
\setstretch{0.1} 
\bibliographystyle{apsrev4-2}
\bibliography{skeleton}
\endgroup

\pagebreak

\section*{Full Author List}

\noindent\includegraphics[width=0.35\linewidth]{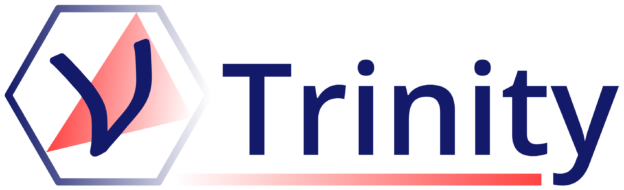}
\vspace{0.5em}

\small
D. R. Bergman$^{2}$,
J. Bogdan$^{1}$,
A. M. Brown$^{3}$,
M. Doro$^{4}$,
M. Fedkevych$^{1}$,
F. Giordano$^{5}$,
C. Hao$^{1}$,
D. Kieda$^{2}$,
M. Mariotti$^{3}$,
Y. Onel$^{6}$,
A. N. Otte$^{1}$,
D. A. Raudales O.$^{1}$,
E. Schapera$^{1}$,
D. Soldin$^{2}$,
W. Springer$^{2}$,
S. Stepanoff$^{1}$,
I. Taboada$^{1}$
K. Tran$^{2}$

\medskip

\noindent
$^{1}$School of Physics and Center for Relativistic Astrophysics, Georgia Institute of Technology, Atlanta, GA, USA \\
$^{2}$Department of Physics and Astronomy, University of Utah, Salt Lake City, UT, USA \\
$^{3}$Department of Physics and Centre for Advanced Instrumentation, Durham University, Durham, UK \\
$^{4}$INFN Sezione di Padova and Università degli Studi di Padova, Padova, Italy \\
$^{5}$INFN Sezione di Bari and Università degli Studi di Bari, Bari, Italy \\
$^{6}$Physics and Astronomy Department, The University of Iowa, Iowa City, USA \\

\end{document}